\documentclass[twocolumn,showpacs,amsmath,a4paper,amssymb,aps,prb,floatfix,groupedaddress]{revtex4}
\usepackage{graphicx} 
\usepackage{dcolumn} 
\usepackage{bm} 
\usepackage{epsfig}
\usepackage{comment}
\usepackage[monochrome]{color}
\usepackage{amsmath}

\begin{document}
\title{Electronic fine structure in the\\
nickel carbide superconductor Th$_{2}$NiC$_{2}$ }

\author{Y. Quan}
\affiliation{Department of Physics, University of California, Davis, CA 95616, USA}

\author{W. E. Pickett}
\affiliation{Department of Physics, University of California, Davis, CA 95616, USA}

\date{\today}

\begin{abstract}
The recently reported nickel carbide superconductor, body centered tetragonal 
$I4/mmm$ Th$_2$NiC$_2$  with T$_c$ = 8.5 K increasing to 11.2 K upon alloying Th with Sc, 
is found to have very fine structure in its electronic spectrum, 
according to density functional
based first principles calculations.
The filled Ni $3d$ band complex is hybridized with C $2p$ and Th character to and
through the Fermi level ($E_f$), and a sharply structured density of states arises only when spin-orbit
coupling is included, which splits a zone-center degeneracy leaving a very flat
band edge lying at the Fermi level.
The flat part of the band corresponds to an effective mass $m^*_{z} \rightarrow
\infty$ with large and negative $m^*_{x}=m^*_{y}$. Although the region over which the 
effective mass characterization applies is less than 1\% of the zone volume, it yet
supplies of the order of half the states at (or just above) the Fermi level.  
The observed increase of T$_c$ by hole-doping is accounted for if the reference
as-synthesized sample is minutely hole-doped, which decreases the Fermi level
density of states and will provide some stabilization. In this scenario, electron doping
will increase the Fermi level density of states and the superconducting critical 
temperature.  Vibrational
properties are presented, and enough coupling to the C-Ni-C stretch mode at 
70 meV is obtained
to imply that superconductivity is electron-phonon mediated. 
\end{abstract}

\maketitle

\section{Introduction}
The discovery of new superconductors in the last dozen years, with MgB$_2$ found
to be superconducting at T$_c$=40 K [\onlinecite{jun2001,isotope}] and 
Fe-based pnictide and chalcogenide
superconductors (FeSCs) up to 56 K, has reinvigorated the search for new
superconductors.  While MgB$_2$ remains a class of one material, the FeSCs like
the cuprates have blossomed into several subclasses with dozens of
members.  The discovery of a new Ni-C superconductor in a distinct structural class
provides reason for rapid follow up to understand its properties, establish its
relationships to other superconductors, and perhaps to extend the class.

The recent discovery by Machado, Grant, and Fisk\cite{machado} (MGF) that Th$_2$NiC$_2$
superconducts at T$_c$=8.5 K, increasing to 11.2 K when alloyed with Sc, raises questions
besides whether it might lead to other members of this structural 
class,\cite{Moss} possibly with higher
T$_c$.  Notably, MGF  have categorized it as a very low density of states (DOS)
superconductor, because the specific heat coefficient extrapolated from a C(T)/T
versus T$^2$ plot is vanishingly small -- certainly an exceptional feature and one that
is difficult to reconcile with metallic and superconducting behavior. Secondly,
it is another transition metal (Ni) based superconductor, calling to mind the
cuprates and FeSCs. 

A central structural motif, Ni bonded to a metalloid (C or B) which in this case
is in the form of the linear C-Ni-C trimer, reminds of
the rare earth (${\cal R}$) Ni borocarbides ${\cal R}$Ni$_2$B$_2$C that have
attracted a great deal of attention,\cite{luni2b2cA,luni2b2cB} largely for the competition between
${\cal R}$ magnetic ordering and superconductivity, and related members. The
compound with the largest T$_c$=23K, YPd$_2$B$_2$C, is based on isovalent Pd,
suggesting substitutions in Th$_2$NiC$_2$. YNi$_2$B$_2$C, and similarly LuNi$_2$B$_2$C,
has a Ni-dominated density of states peak\cite{mattheiss,luni2b2cC} at E$_F$ that 
promotes superconductivity (as well as other instabilities). These compounds have a B-Ni-B stretch
mode originally suspected to be strongly coupled to Fermi surface (FS) 
electrons,\cite{luni2b2cC,luni2b2cD}
but full calculations indicated coupling spread over much of the vibrational spectrum.\cite{yni2b2cA} 

Other Ni-based superconductors, often with C neighboring Ni, have
appeared in other crystal classes.  Anti-perovskite MgCNi$_{3}$\cite{he2001} with T$_c$ =
8 K occurs with E$_F$ lying in a narrow and extremely sharp van Hove singularity,\cite{rosner}
where superconductivity competes strongly with ferromagnetic tendencies in a manner
that is not typical of other Ni-based superconductors.  Heusler compounds
such as ZrNi$_2$Ca which contain no C or B metalloid atom, and
which are more often magnetically ordered when
based on $3d$ transition metals, superconduct typically around T$_c \sim$ 3K.\cite{winterlik}
The quaternary LaNiBN, where Ni is coordinated with B separated by LaN layers, superconducts at 4 K [\onlinecite{lanibn1,lanibn2}] and its Pt analog\cite{laptbn} does so at 6K. 
This compound is structurally and electronically\cite{la3ni2b2n3} related to La$_3$Ni$_2$B$_2$N$_3$, 
also with Ni coordinated with B and superconducting at 12 K.
More closely related to the compound of interest here, Th$_2$NiC$_2$, are two other
compounds discovered by Fisk and collaborators, Th$_3$Ni$_5$C$_5$ with\cite{th3ni5c5} T$_c$=5 K,
and the Ni-rich member of this ternary system ThNi$_4$C, with\cite{thni4c} T$_c$=5.5 K.
A ``nickel carbide'' LaNiC$_2$ superconducts at 2.7 K and electron-phonon coupling is
responsible,\cite{lanic2} but it is based structurally on Ni-Ni and C-C dimers 
within a La framework rather
than on Ni-C bonding.

Information so far\cite{machado} points to the stoichiometric compound Th$_2$NiC$_2$ being
a thermodynamically stable material only at high temperature.  When alloyed on the tetravalent Th
site with trivalent Sc, the material becomes stable in air and achieves the 11.2 K
superconductivity mentioned above.  The normal state resistivity and heat capacity
follow conventional forms, and the extrapolated upper critical field is H$_{c2}
\sim 11-12$ T. MGF suggested 
neutral Ni $d^{10}$ to be the appropriate picture, which we will 
clarify in our study of the electronic structure.
The apparently vanishingly small DOS at the Fermi level is a central point of
interest in this system.

In this paper we present density functional theory (DFT) based results for the electronic properties and
vibrational structure of Th$_2$NiC$_2$ at stoichiometry and when alloyed with Sc.
The electronic structure is representative of what should be treated accurately from
the DFT viewpoint -- the Ni electronic structure is similar to a few other Ni-C and Ni-B
compounds with filled $3d$ bands that
are described well by such calculations. Shein and Ivanovskii (SI) have provided some
results\cite{Shein} of electronic structure calculations but focused primarily on structural
properties. We find that while Th$_2$NiC$_2$ is a relatively low N(E$_F$) metal, 
the extremely low (essentially vanishing) value inferred by MGF from extrapolation of heat 
capacity data
from above T$_c$ seems inconsistent with DFT results. Unusually fine structure in N(E)
is found to lie precisely at the Fermi level, providing a resolution to some of the questions
about this system and suggesting unconventional behavior, {\it viz.} thermodynamic properties.

The organization of this paper is as follows. After describing the methods and crystal
structure in Sec. II, the electronic structure of Th$_{2}$NiC$_{2}$ is
analyzed and contrasted with related Ni-C compounds in section III.
Section IV deals with the electronic structure of doped Th$_{2-x}$Sc$_x$NiC$_{2}$.
Phonon dispersion curves of Th$_{2}$NiC$_{2}$ are presented in Sec. V, and we look at how the C-Ni-C bond stretch mode affects the electronic structure in Sec. VI, to obtain insight into possible strong coupling of this mode. In section VII, the impact of band anomalies is discussed. We summarize our results in the final section.

\section{Crystal Structure and Theoretical Methods}
\begin{figure}[!htb]
\begin{center}
\includegraphics[width=0.4\textwidth,angle=0]{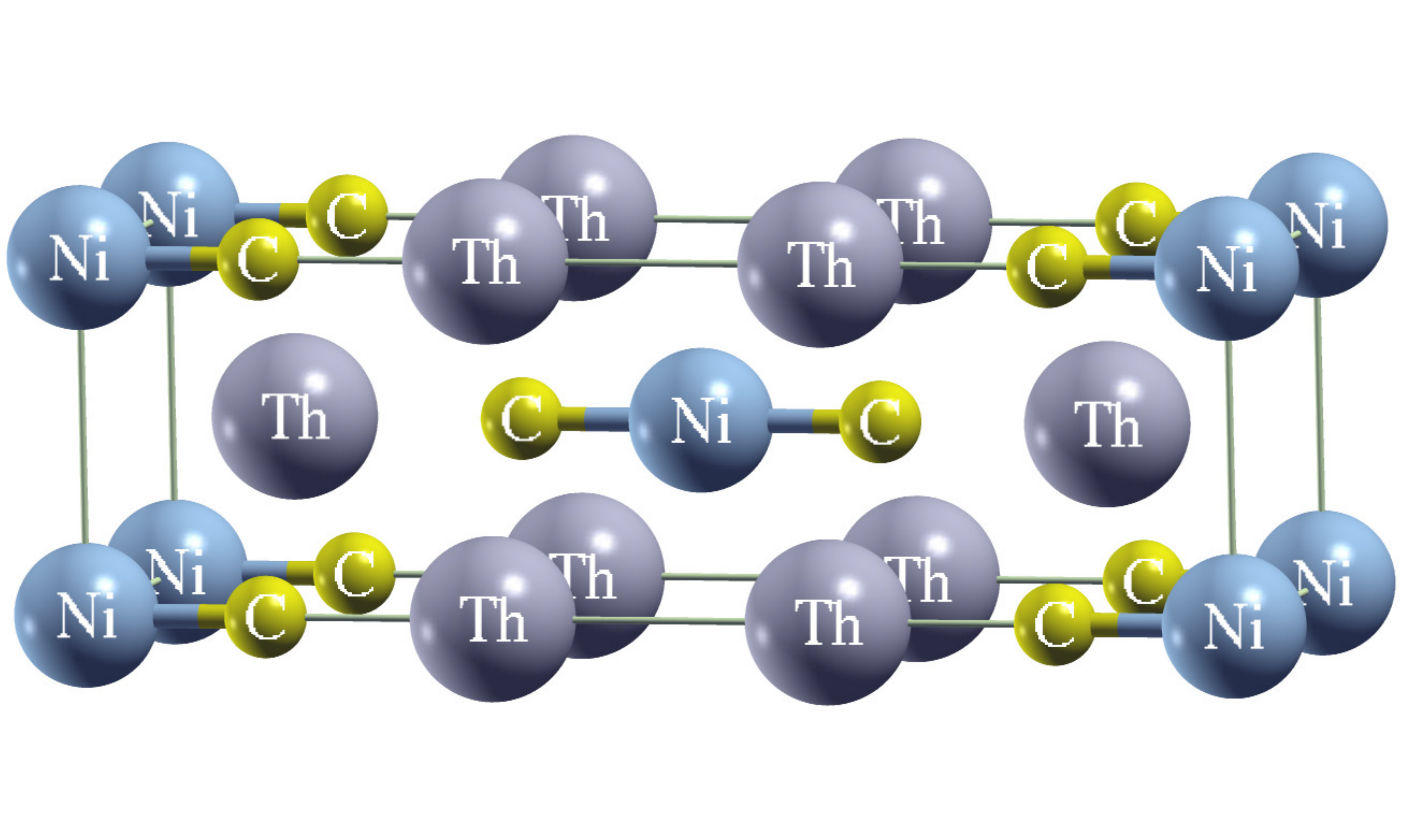}
\end{center}
\caption{(Color online)
The $bct$ $I4/mmm$ U$_2$IrC$_2$ crystal structure type of Th$_2$NiC$_2$.
The $c$-axis is horizontal. The C-Ni-C trimer appears
at the center of this figure; Th-C bonds discussed in the text are not
shown. The structure can be pictured as a $bct$
array of linear Th-C-Ni-C-Th units, with  neighboring units
containing nearly co-planar layers of Th and C atoms.  The
short Ni-C distance is a noteworthy aspect of the structure, but the Th-C
bonds display covalent bonding aspects as well.\cite{xcrysden}
\label{fig:Th2NiC2structure}
}
\end{figure}

The crystal structure is taken from Moss and Jeitschko\cite{Moss} and is shown in
Fig. ~\ref{fig:Th2NiC2structure}. The space group of Th$_2$NiC$_2$ is tetragonal
$I4/mmm$ (\#139), with five atoms per primitive cell and lattice constants
$a$=3.758\text{AA}, $c$=12.356\AA. The Wyckoff positions are
$4e$ (Th, C) and $2a$ (Ni),
with site symmetries $4mm$ and $4/mmm$ respectively. The internal coordinates
from
single crystal X-ray data are $z_{Th}$ = 0.354, $z_C$ = 0.156. The nearest neighbor
distances are Ni-C 1.93$\text{\AA}$, Th-C 2.45$\text{\AA}$ and 2.66$\text{\AA}$, and Th-Ni 3.21$\text{\AA}$.
We will find it useful to consider the structure as a body centered tetragonal
array of linear Th-C-Ni-C-Th units oriented along the $c$-axis.  SI have provided
evidence from charge density contours of directional (presumably partially
covalent) Th-C and C-Ni bonding.

The internal structure of Th$_{2}$NiC$_{2}$ was relaxed, keeping the lattice constants at their
experimental values. The calculated relaxed values $z^c_{Th}$=0.3536, 
$z^c_C$=0.1545 are very close to the
observed values: that of Th is indistinguishable from the experimental value to the precision quoted for
the data, and the difference of 0.0015$c$=0.017\AA for the C position is small. 

The linearized augmented plane wave code $WIEN2k\_11$\cite{wien} has been used unless noted
otherwise. All calculations are carried out using local density approximation (LDA).
The muffin-tin radii of Th, Ni and C are 2.44 $a_0$, 1.92 $a_0$ and 1.70 $a_0$
respectively. K-space is sampled over a fine mesh of 3349 special k points
in the irreducible wedge of the Brillouin zone.
RK$_{max}$ is chosen to be 8, making the plane wave cutoff  22.15 Ry.
These values provide excellent convergence of the presented results.

Magnetic order was not obtained in the local spin density approximation,
in agreement with observations and as expected from filled Ni $3d$ bands.  As an
actinide with nuclear charge Z=90, spin-orbit coupling must be included for Th
and it has been included for Ni and C as well.
The Th $5f$ bands lie
4 eV above the Fermi level
E$_F$, but significant $5f$ character adds to the $6d$ and $7p$ Th character
due to band mixing.
\noindent
\section{Electronic structure of Th$_{2}$NiC$_{2}$}
\begin{figure}[!htb]
\includegraphics[width=0.32\textwidth,angle=90]{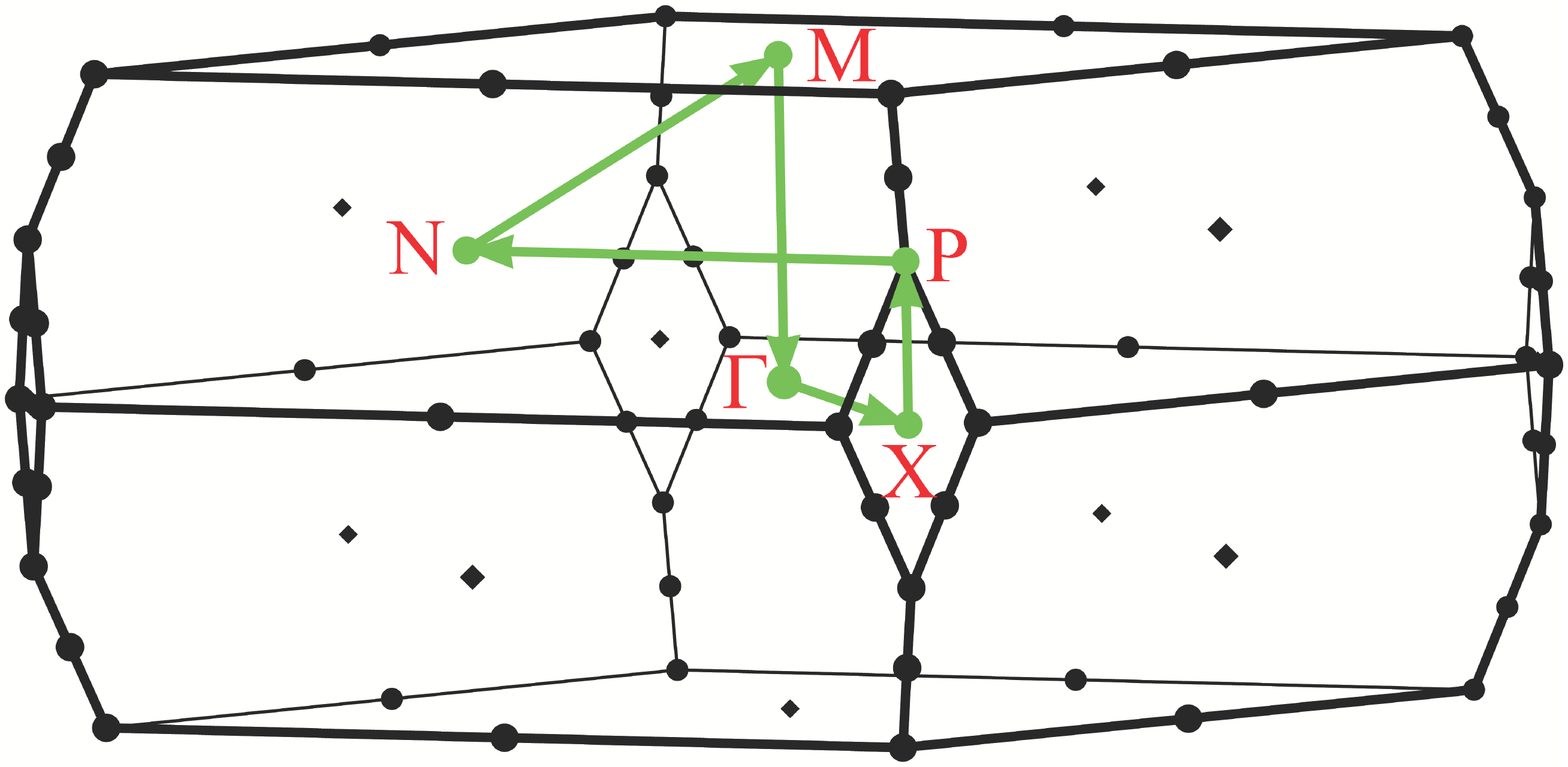}\\
\vskip 8mm
\includegraphics[width=0.5\textwidth,angle=0]{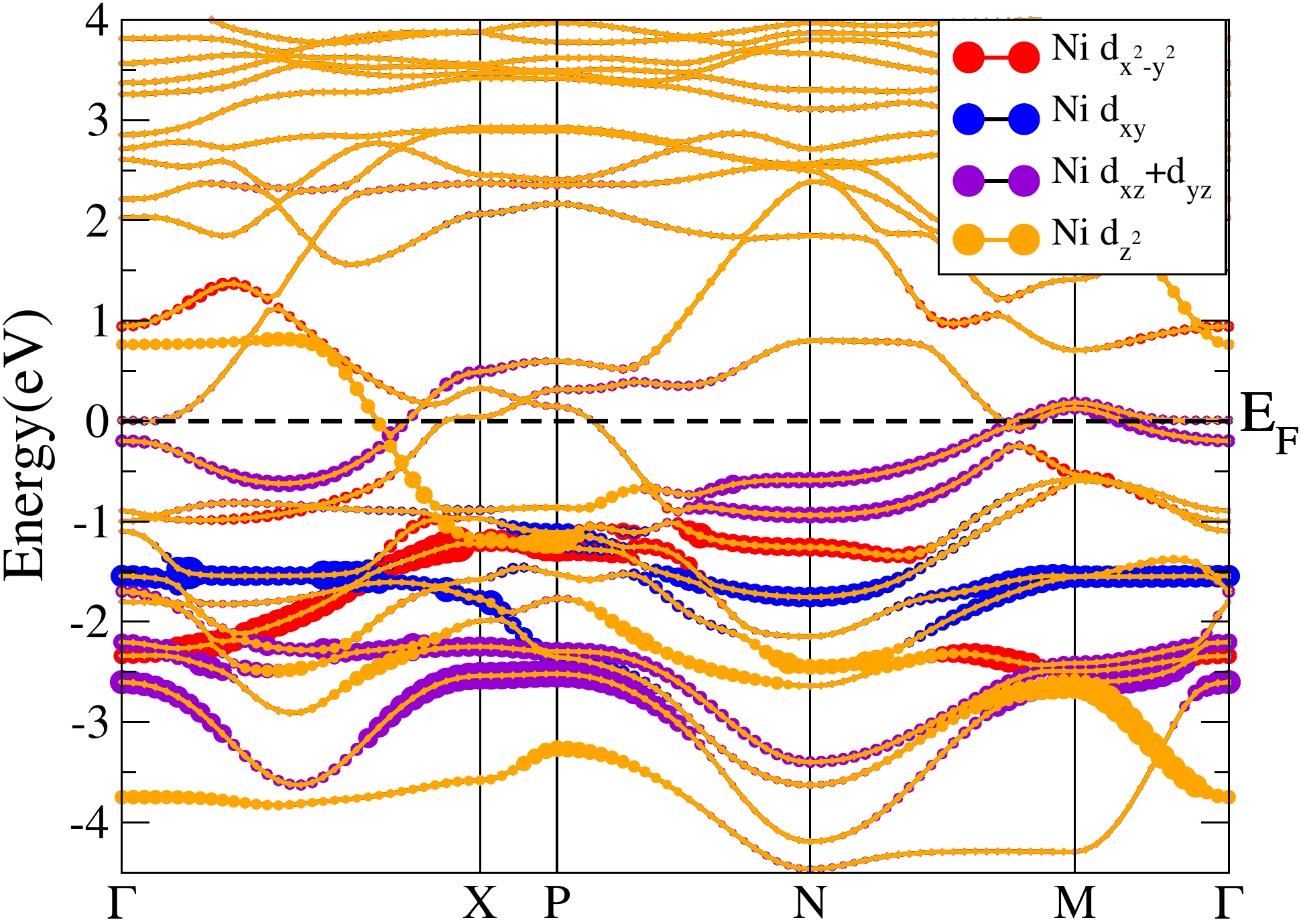}\\
\vskip 8mm
\includegraphics[width=0.5\textwidth,height=1.8in,angle=0]{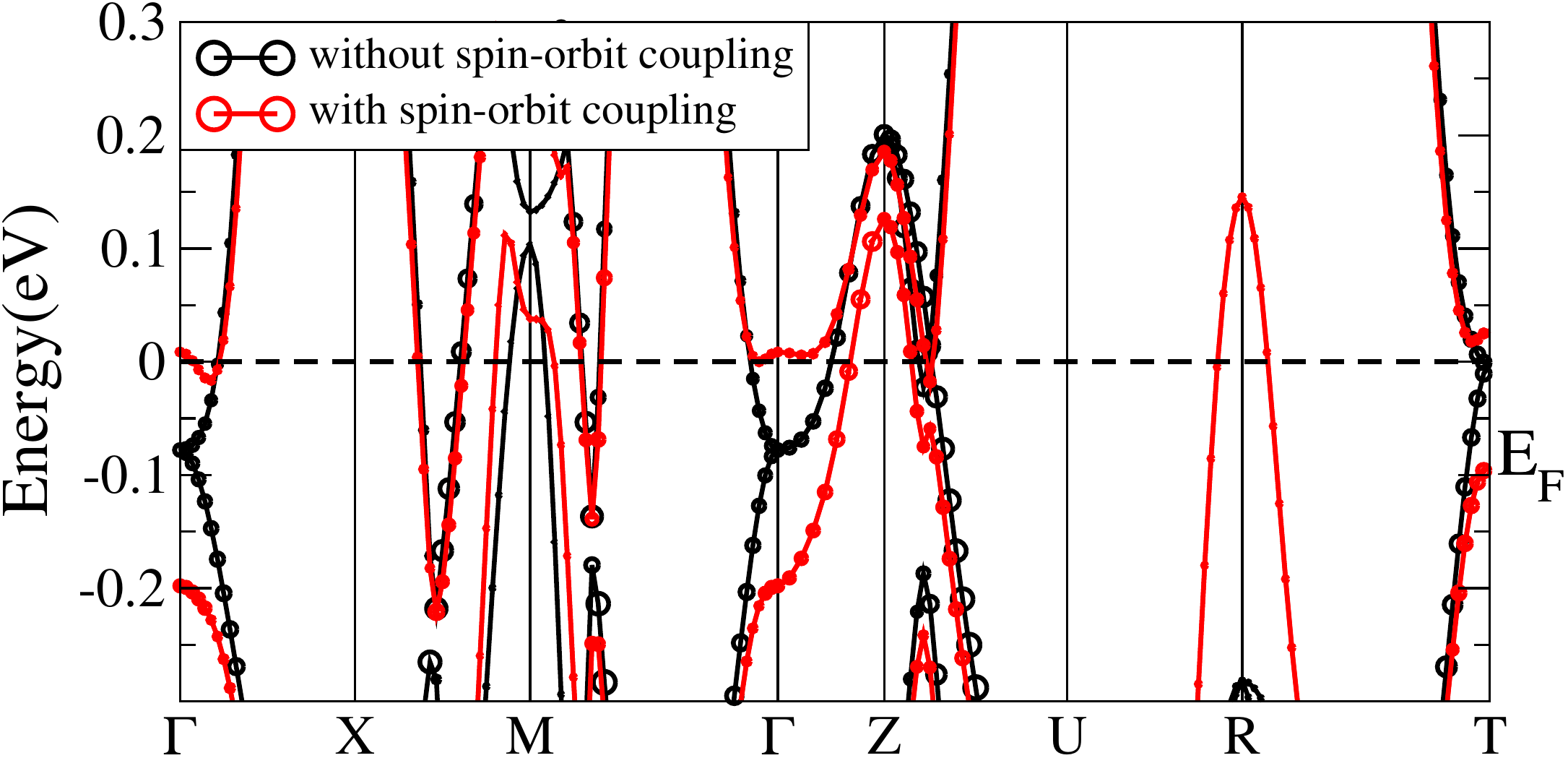}\\
\caption{(color online) 
Top panel: bct Brillouin zone with paths used for the band plot in the middle panel. 
Middle panel: Th$_{2}$NiC$_{2}$ band structure with spin-orbit coupling included,
for the experimental structure, with Ni
$3d$ fatbands shown. 
Bottom panel: expanded region near the Fermi level, illustrating how spin-orbit
coupling (red bands) splits a degeneracy at $\Gamma$ and creates a very flat band 
at the Fermi level (E=0) centered
at $\Gamma$. Bands without SOC are in black.  A large
effect of spin-orbit coupling also arises at the zone corner R=($\pi,\pi,\pi)$.
This panel uses Cartesian lines: X=($\pi,0,0$), M=($\pi,\pi,0$), Z=($0,0,2\pi$),
U=($\pi,0,\pi$), T=($0, 0, \pi$).
}
\label{Bands_Ni3d}
\end{figure}

The band structure along bct symmetry lines, with Ni $3d$ orbital character
emphasized by the fatband representation, is pictured in Fig. \ref{Bands_Ni3d}. In this section
the experimental structure is used except where noted. The
Ni $3d$ character lies in the -3.5 eV to 0.5 eV range; the Ni $3d$ bands are
effectively filled, with some character extending to and above E$_F$ due to
hybridization.  Bands crossing E$_F$ along
N-M-$\Gamma$ have strong
Ni $d_{xz}$,$d_{yz}$ character; other bands have stronger C $2p$ character. 
The Ni $d_{xy}$ weight lies in
the -2 eV to -1 eV range, largely
concentrated in a flat band around -1.5 eV.
 
The dominant feature of the band structure is an extraordinarily flat band
at and near $k$=0. The band can be characterized by an effective mass 
$m^*_{z}\rightarrow \infty$ ({\it i.e.} flat), 
with $m^*_x = m^*_y$ being very large and negative.
This description holds only for a small region, as seen in the lower panel
of Fig. \ref{Bands_Ni3d}: along $k_x$ to about 20\% of
the zone boundary; along $k_z$ to roughly 30\% of the zone boundary.  This
region comprises about 1\% of the zone volume, {\it i.e.} 0.01 states/spin. The 
bottom panel of Fig. \ref{Bands_Ni3d} demonstrates that the flat band
arises due to SOC that splits the degeneracy of a band at -0.1 eV, pushing
the top member of the pair precisely (and accidentally) to E$_F$.  
A band is also driven by SOC very near E$_F$ at the zone corner. SOC expands the FS near the corner by pushing a band above E$_F$ at R. This Fermi level fine structure is reminiscent of the Heusler superconductor (T$_c$=2.9 K) ZrNi$_2$Ga where the Fermi level lies near
an electronic van Hove singularity.\cite{zrni2ga}

\noindent
\begin{figure}[!htb]
\begin{center}
\includegraphics[width=0.45\textwidth,angle=0]{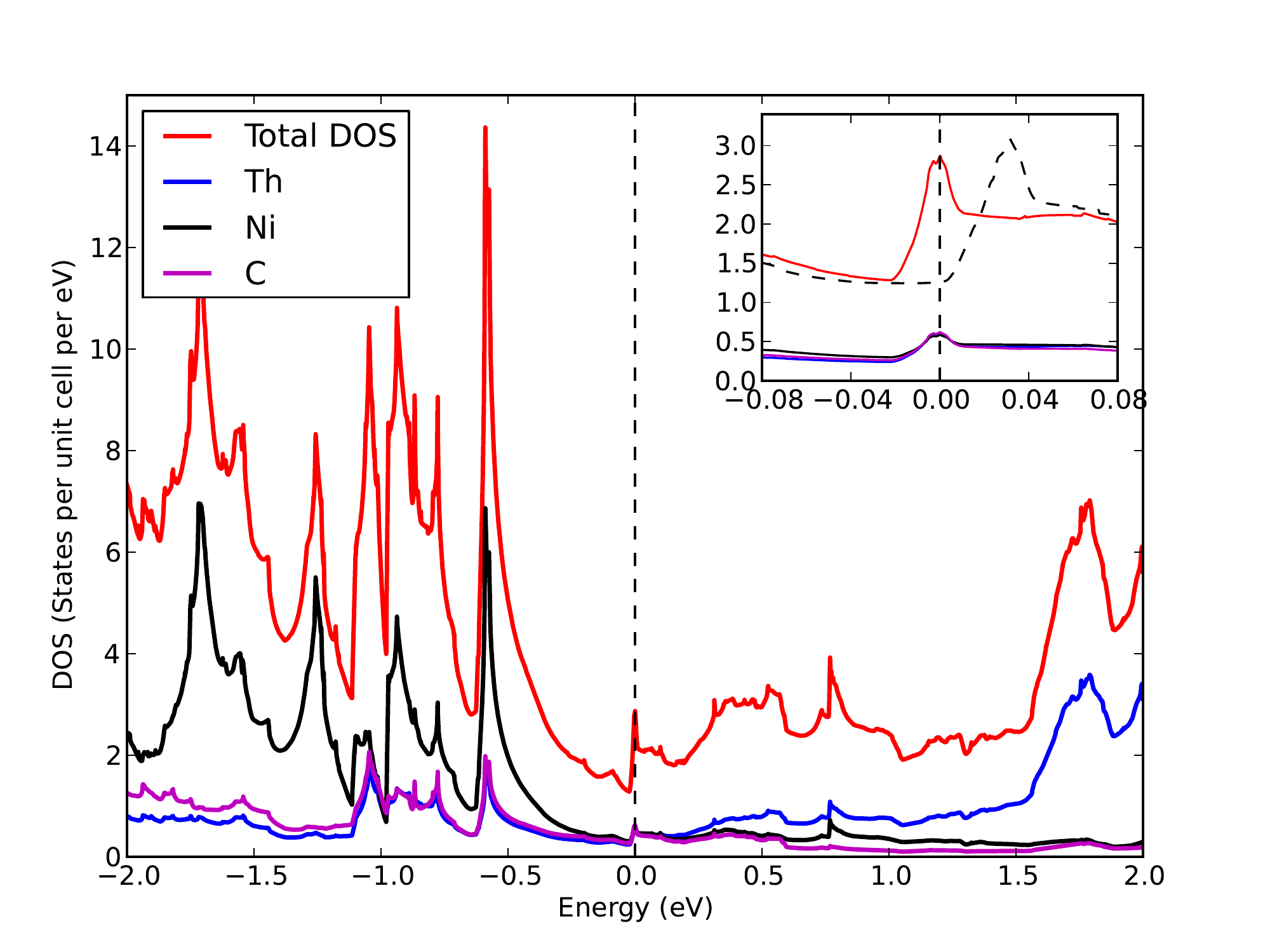}
\includegraphics[width=0.45\textwidth,angle=0]{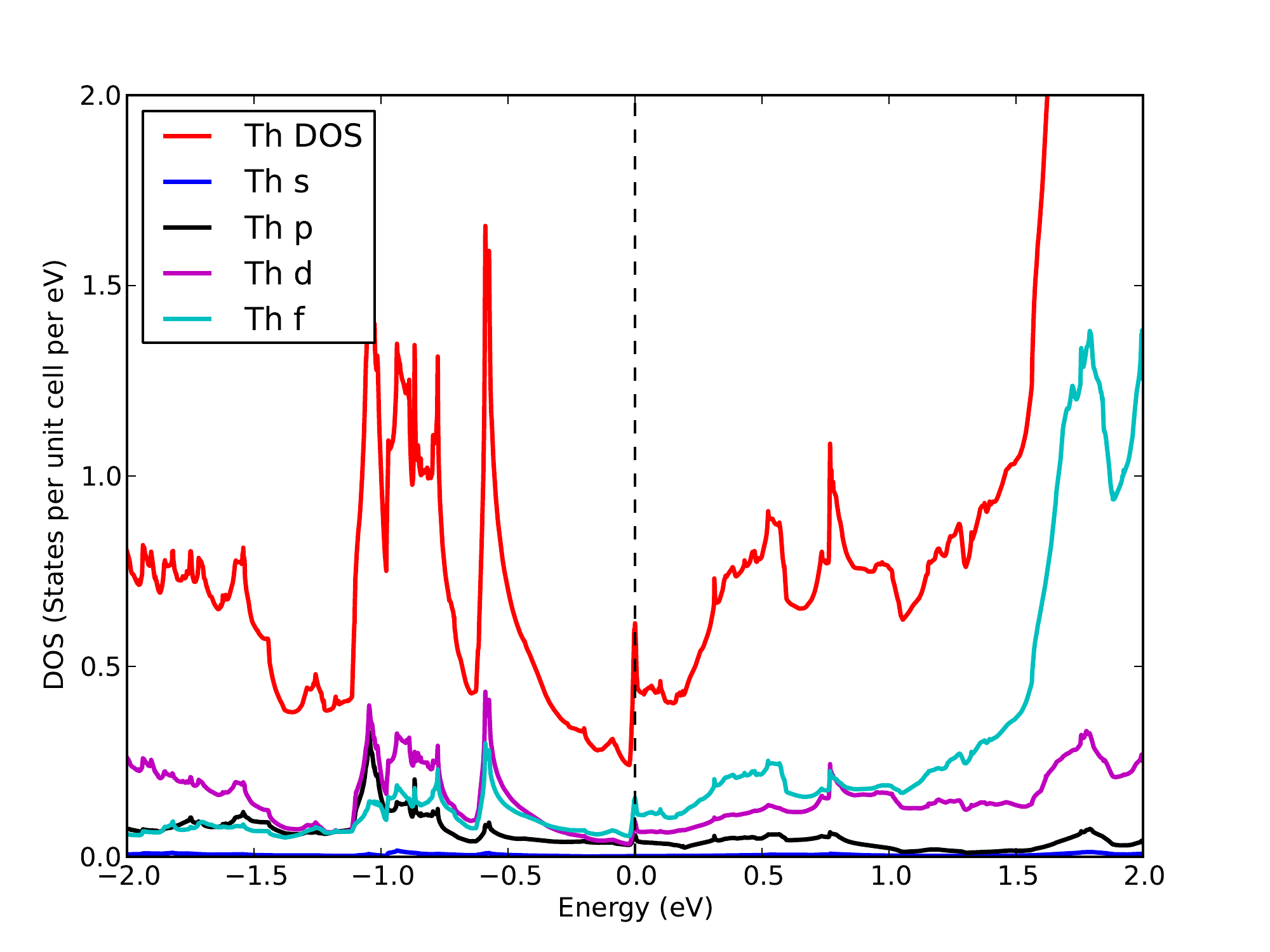}
\includegraphics[width=0.45\textwidth,angle=0]{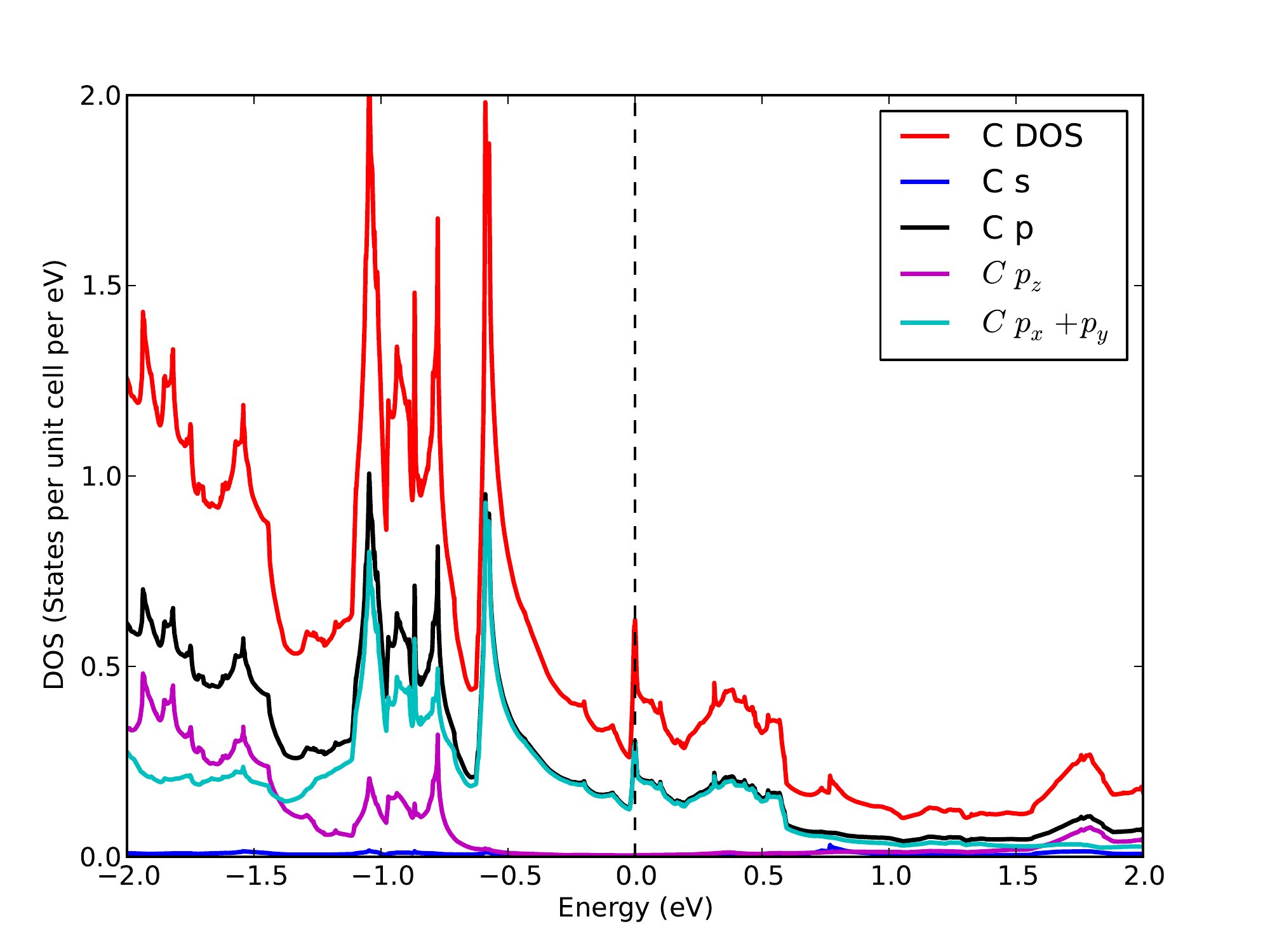}
\end{center}
\caption{(Color online) {Total and orbital-projected densities of states for 
Th$_2$NiC$_2$. Top panel: total DOS
and the Ni $3d$ contribution. The DOS has very sharp structure
at the Fermi level, as shown in the inset on an expanded scale. The dashed curve in
the inset shows the fine structure using the relaxed atomic positions.
Middle panel: Th $s, p, d, f$ contributions; $f$ and $d$
contributions are dominant at E$_F$.
Bottom panel:
C $2s$, $2p$ projected DOS, the negligible $2s$ contribution
is hardly visible, and $p_z$ is negligible at E$_F$. 
}
\label{fig:th2nic2th}
}
\end{figure}

The densities of states (DOS), total and atom-projected, are provided in Fig. \ref{fig:th2nic2th}.
The total N(E$_F$) = 2.8 states/eV per formula
unit arises almost equally from each of the atoms.
E$_F$ lies at the center of a very narrow, $\sim$30 meV wide peak shown in the inset,
that arises from the flat band
discussed above. Without looking at the fine scale, the Fermi level appears to lie on
the edge of a step discontinuity of the DOS. The flat band has similar size 
contributions from Ni (mostly $d_{xz},d_{yz}$), C ($p_x,p_y$), and Th (all of $5f$,$6d$,$7p$).
This fine structure will have impact on the
measured properties of Th$_2$NiC$_2$, a topic we return to later. 

\begin{figure}[!htb]
\begin{center}
\includegraphics[width=0.5\textwidth,angle=0]{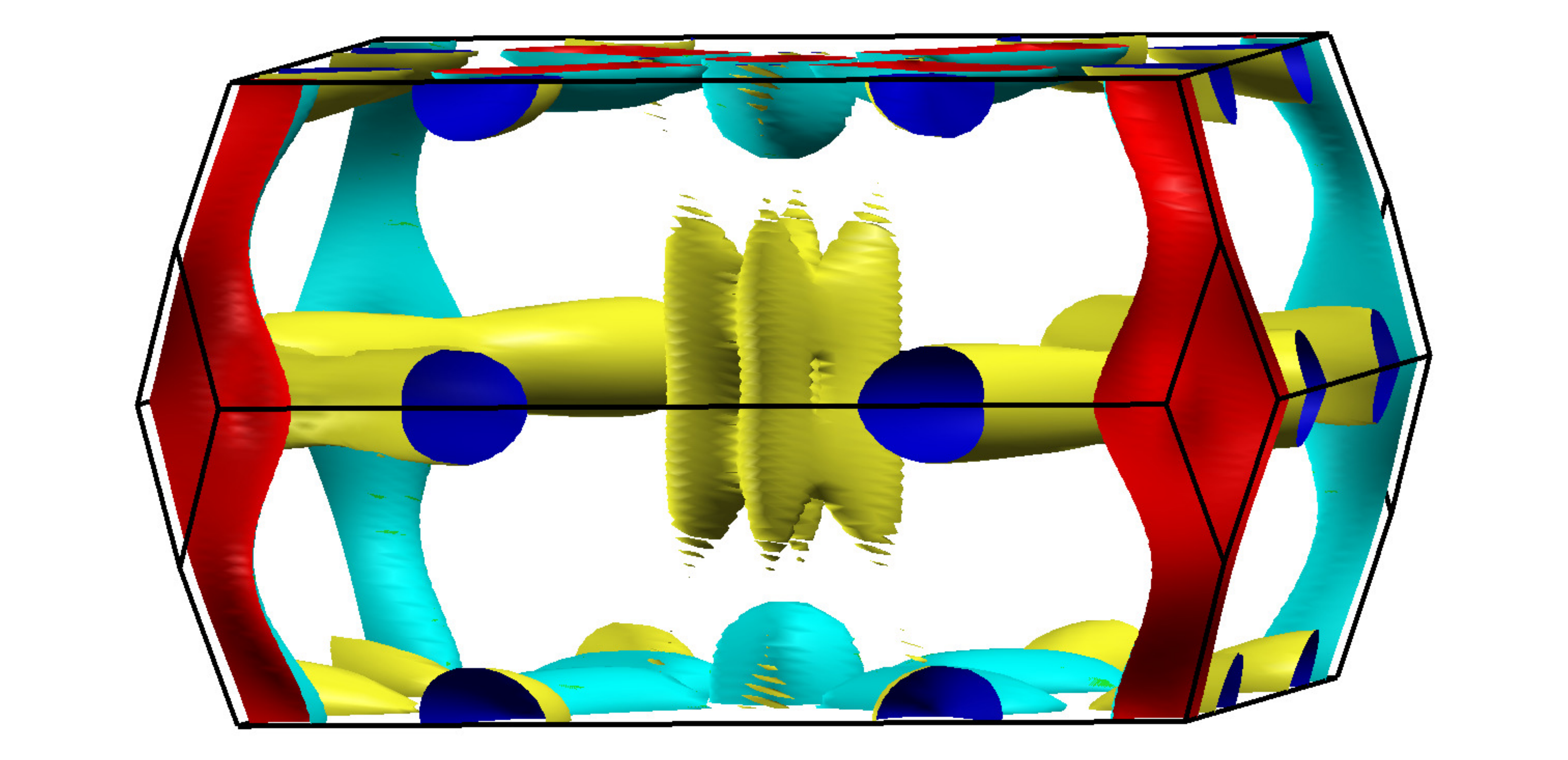}\\
\end{center}
\caption{(Color online) Fermi surfaces of Th$_2$NiC$_2$, as described in the text. The ripples
of the $\Gamma$-centered sheet arise from difficult (hence inaccurate) 
numerical interpolation in the region
of the kink in the band at the Fermi level. 
\label{fig:fermisurface}
}
\end{figure}

This fine structure is
somewhat sensitive to the internal structural parameters.   
Also shown in the inset is N(E) when atomic positions are optimized to minimize the energy.
Though the difference in atomic positions is quite small (to the accuracy provided by the 
experimental data the C position, and Ni-C distance, are indistinguishable), 
the structure is so sharp and so close
to E$_F$ that N(E$_F$) is strongly affected.  
The FS of Th$_2$NiC$_2$, shown in Fig ~\ref{fig:fermisurface}, is connected along $z$ axis
by tubes along the zone corners. Another surface (``tube'') lies in the $k_x-k_y$ plane, which
becomes the zone top (or bottom) of the next BZ.  Centered at $\Gamma$ is a fourfold
closed surface of connected ``cigars'. Fermi surface near $\Gamma$ are determined primarily 
by the flat band discussed above.
Overall, the Fermi surface is representative of a
good, not unusually low DOS, metal.

\noindent

\section{Hole doping on the {\rm Th} site}
\begin{figure}[!htb]
\noindent
\begin{center}
\includegraphics[width=0.40\textwidth,angle=0]{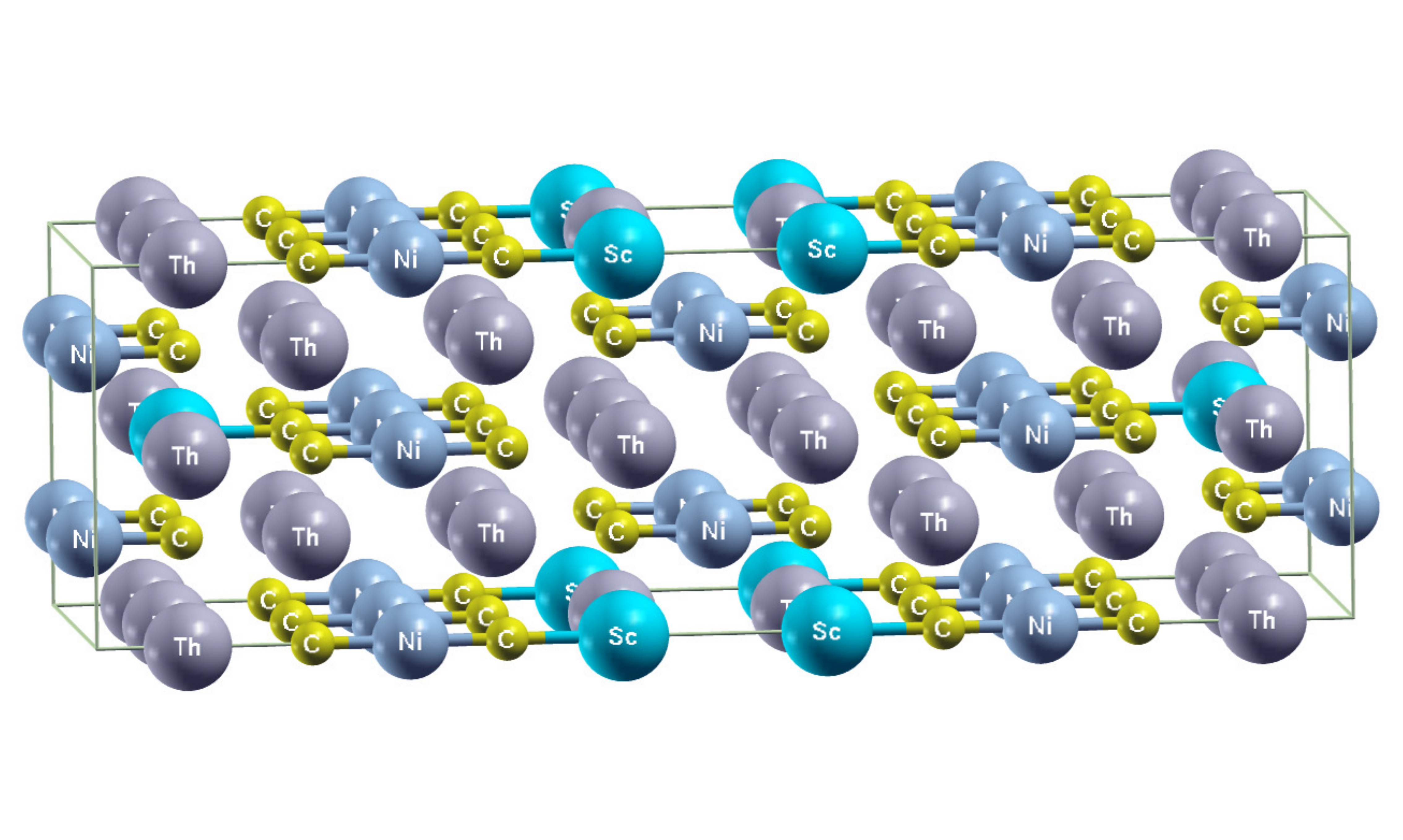}\\
\end{center}
\caption{(Color online) Crystal structure of the supercell calculation. Sc dopant atoms
are 2nd neighbors on the Th sublattice in the $x-y$ plane (which is vertical in
this plot).}
\label{fig:supercellstructure}
\end{figure}

\begin{figure}[!htb]
\noindent
\begin{center}
\includegraphics[width=0.48\textwidth,angle=0]{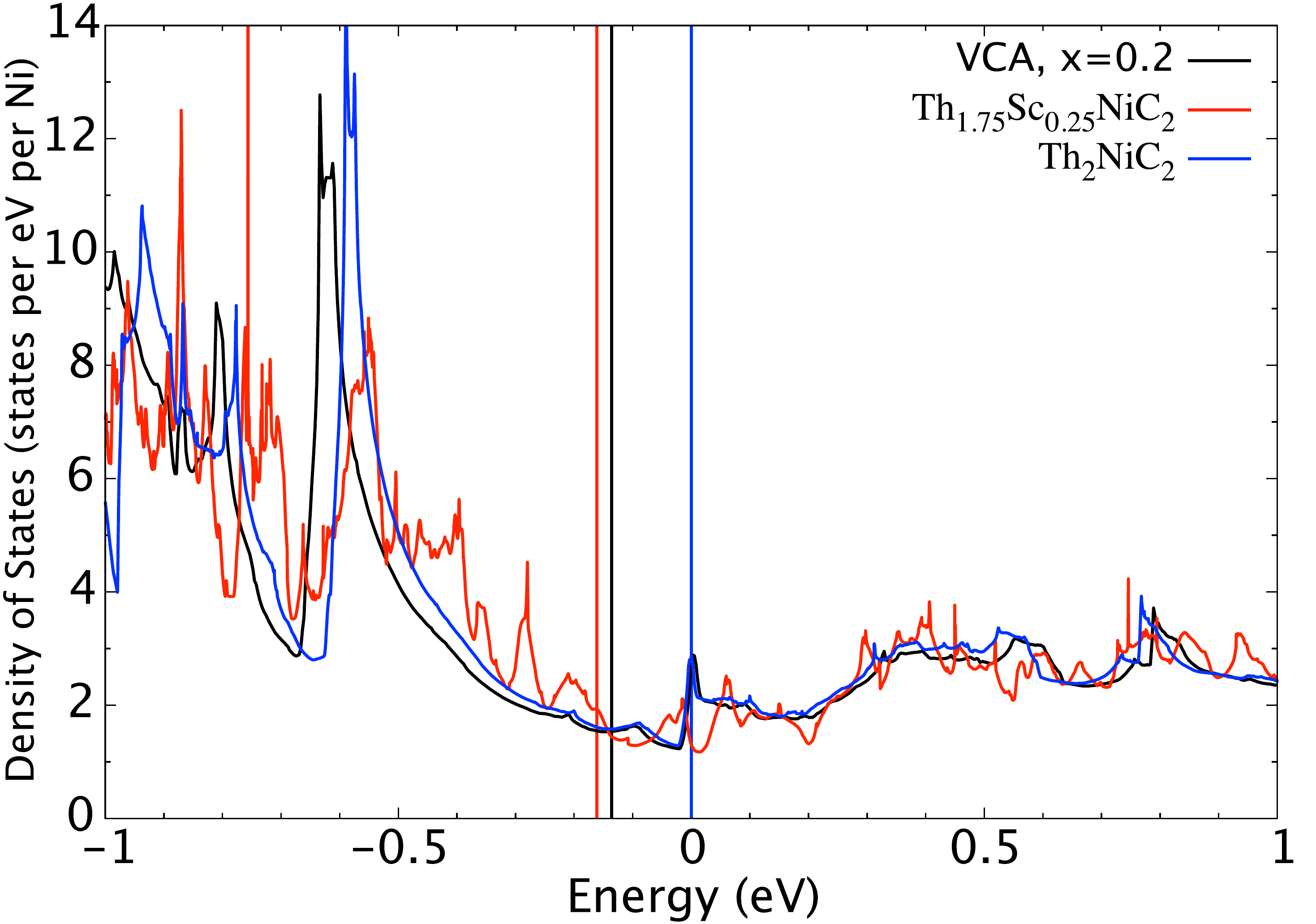}\\
\end{center}
\caption{(Color online) DOS for stoichiometric Th$_2$NiC$_2$ and two types of 
treatment of doping:
virtual crystal (VCA) ``Th$_{1.8}$Sc$_{0.2}$NiC$_2$'' 
and ordered alloy Th$_{1.75}$Sc$_{0.25}$NiC$_2$
$\equiv$ Th$_7$ScNi$_4$C$_8$.
The Fermi level of stoichiometric Th$_2$NiC$_2$ is chosen as the common energy zero, with the alloy results
aligned at the same band filling.  The Fermi level drops with doping as denoted by the
horizontal lines. The ordered alloy DOS contains small sharp structure that will be
broadened by random Sc substitution. See text for discussion of redistribution of spectral density
due to Sc doping.}
\label{fig:dopedDOS}
\end{figure}

Doping with Sc on the Th site increases the
superconducting T$_c$ by a few Kelvins ($\sim$30\%).
SI suggested that doping with Sc leads to larger density of states
at the Fermi level,\cite{Shein} which could explain the increase in T$_c$.
We have carried out two types of calculations, both including SOC and
including effects beyond the rigid band behavior assumed by SI.
In the first calculation, a 2 $\times$ 2 $\times$ 2 
supercell of the conventional cell is
adopted. Four of the 32 Th atoms were substituted with Sc, giving ordered
Th$_{1.75}$Sc$_{0.25}$NiC$_{2}$ (12.5\% doping).
The Sc positions were chosen as shown in Fig. ~\ref{fig:supercellstructure}. The small 
decrease in cell volume observed in experiment was neglected, and R$_{mt}$K$_{max}$
was reduced to 7 which is sufficient for this system. After relaxation of the atomic 
positions in this supercell,
there are small variations of the inequivalent Ni-C bond lengths: 1.92$\text{\AA}$, 1.89$\text{\AA}$ and 
1.90$\text{\AA}$. Sc-C distances are 2.29 $\text{\AA}$ (along c direction) and 2.60$\text{\AA}$ (almost within 
the $a-b$ plane.) In the pure Th$_2$NiC$_2$ compound, the Th-C bond length is 2.45 $\text{\AA}$ 
along the $z$ direction and 2.66 $\text{\AA}$ (almost) within $a-b$ plane. The largest difference
thus is in the Sc-C (versus Th-C) separation.
In the second calculation the virtual crystal approximation (VCA) was adopted, with 10\% hole-doping
being modeled by decreasing the nuclear charge of Th to 89.9, which corresponds in
terms of valence electron count to
Th$_{1.8}$Sc$_{0.2}$NiC$_2$. VCA treats the change in electron count self-consistently, hence it
is more realistic than simple rigid band, but it neglects chemical and size differences. 
Treating doping within VCA can however lead to substantial changes in the electronic structure,
as shown recently for the case\cite{cualo2} of hole-doping in CuAlO$_2$. 
In that compound, substituting Mg on the Al site
decreases the transfer of charge from the cation (Al, Mg) to Cu, with a resulting 
dipole-layer shift in potential. In that compound the electronic structure changes
dramatically, with a strong decrease in the strength of electron-phonon coupling.\\
 Because N(E) in Th$_2$NiC$_2$ near E$_F$ arises equally from the three atoms, the change in valence
electron density due to doping will be spread evenly throughout the cell, and indeed we
find relatively minor differences compared to the rigid band picture.
Comparison of N(E) for the three calculations is provided in Fig. ~\ref{fig:dopedDOS}.
To better assess the effect of doping on the spectrum,
the density of states of both alloy models was shifted for display so that the total number of
states per unit cell below E=0 are the same.
The virtual crystal
calculation results in rigid band behavior down to and somewhat beyond the 
Fermi level of the doped material.
Due to the DOS fine structure, doping by the first $\sim$0.01 holes may decrease
N(E$_F$) by up to a factor of two; as pointed out in Sec. III, the fine structure and thus
N(E$_F$) is sensitive to the internal parameters. Beyond the initial drop,
N(E) recovers slowly and non-monotonically with further hole doping. \\
For the supercell with 12.5\% ordered Sc substitutionals,
the profusion of bands (2$\times$2$\times$2 times as many) with their anticrossings give
rise to numerous additional van Hove singularities in N(E) because the supercell has
16 times as many bands as the original compound.  This complexity is configuration
specific, and the DOS will be smeared in 
the disordered alloy. Without knowledge of the amount of smearing, which would
require a coherent potential approximation alloy calculation to reveal, spectral
shifts cannot be determined with certainty.  However, the decrease in the strength
of the peak near -0.6 eV and increase in the DOS in the -0.5 eV to -0.15 eV in
the supercell calculation 
suggests that Sc doping does lead to some general increase in spectral density 
in this region beyond what
the VCA picture gives.\\
The very fine structure precisely at E$_F$ for the stoichiometric compound 
suggests a scenario that may account for the increased
T$_c$ with hole doping.  Ternary compounds can be difficult to prepare in absolutely
stoichiometric form, and the difficulty is increased for marginally stable
materials.  We suspect therefore that Th$_2$NiC$_2$ as synthesized is slightly
hole doped; this will decrease N(E$_F$) by up to a factor of two and increase
its stability due to the reduction of the charge and spin susceptibilities.  
With respect to this ``undoped'' reference material, Sc doping
by 5-10\% should indeed increase  N(E$_F$) somewhat and therefore increase T$_c$,
more or less as observed.   
\section{Phonon Dispersion}
Based on experience from several other intermetallic Ni-based superconductors, 
we pursue the likelihood that the superconducting
mechanism is electron-phonon coupling, so assessment of the phonon dispersion curves is needed.
The code {\it phonopy}\cite{phonopy} was used. For this, forces calculated from
finite displacements of atoms in a 3$\times$3$\times$1
supercell are used to compute the elastic constants which can be used to calculate the dynamic 
matrix at an arbitrary q-point.  The relaxed atomic positions were used as the reference structure;
as noted in Sec. II these positions are close to the experimental values.
Kohn anomalies, which are not evident here, would be
reproduced in a smoothed fashion with this method, since they involve very long-range force constants.

\begin{figure}[!htb]
\noindent
\begin{center}
\includegraphics[width=0.42\textwidth,angle=0]{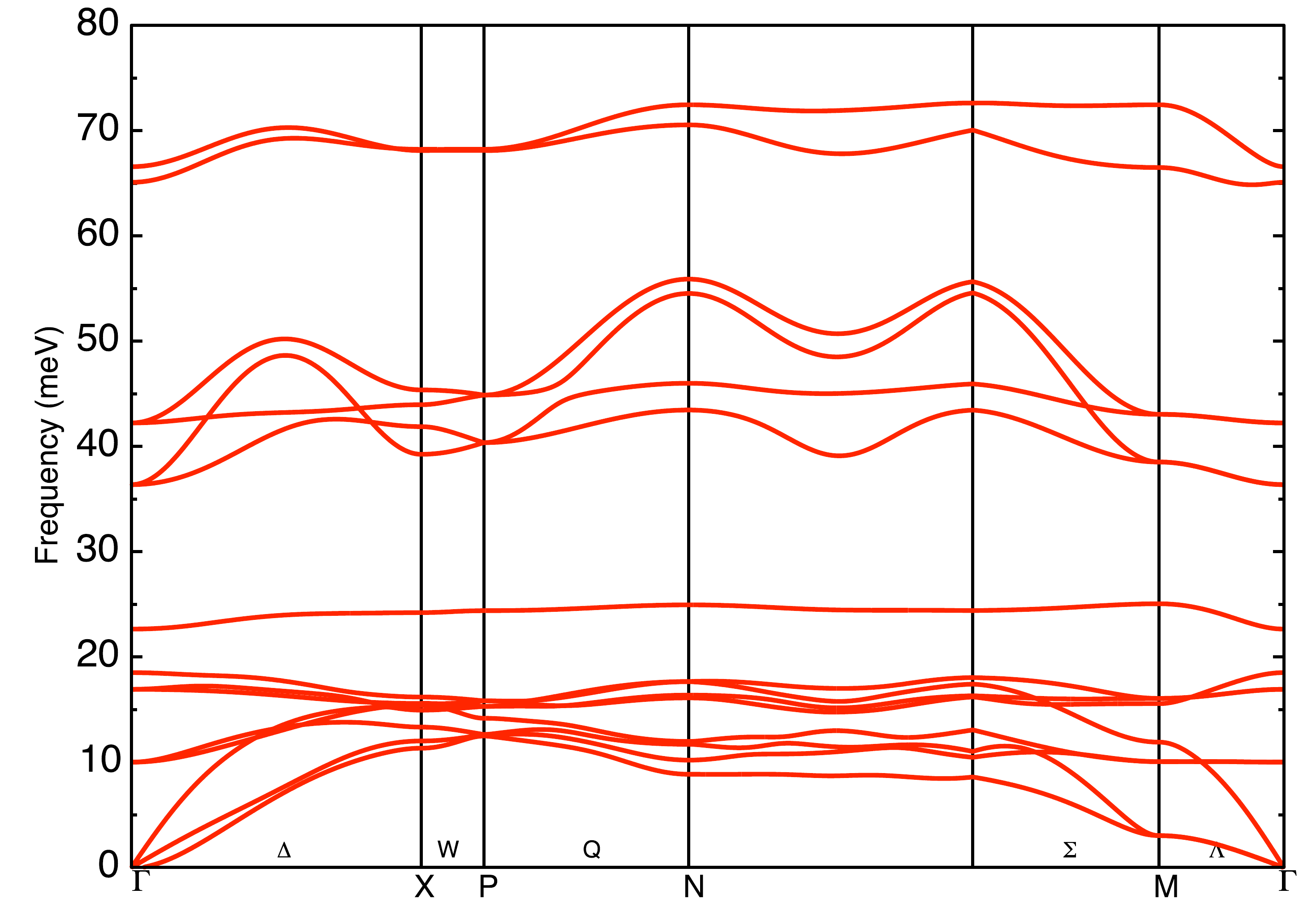}\\
\end{center}
\caption{(Color online) 
Dispersion of phonon branches along this symmetry line path shown in Fig. 2a,
for stoichiometric Th$_2$NiC$_2$. All branches are 
relatively dispersionless except for the necessarily linear acoustic modes. The phonon 
frequency is expressed in units of meV (1THz=4.14meV)}
\label{fig:Th2NiC2phonon}
\end{figure}

The resulting dispersion along bct symmetry lines
is displayed in Fig. ~\ref{fig:Th2NiC2phonon} and
mostly show weak dispersion throughout the zone.
It is worthwhile to distinguish the four (of the fifteen) branches with polarization along $\hat z$,
and consider them in terms of the linear Th-C-Ni-C-Ni unit.  There are two symmetric, Raman-active
A$_{1g}$ modes in which Ni is static; the symmetric C and Th motions can be either in-phase or out-of-phase.
The odd, IR-active A$_{1u}$ modes involve displaced Ni (along, say, +$\hat z$), with one of the
C and Th displacement in-phase and one out-of-phase (the center of mass must be stationary).

The phonon branches separate into four distinct energy regions.\\
$\bullet$
The eight branches below 18 meV include the three acoustic branches, which contain contributions from all
atoms. This complex also contains two pairs of degenerate optic modes, one comprised primarily of
Th and C motions along $\pm\hat x$ and (degenerately) along $\pm\hat y$, and the other pair involving all three atoms moving in the x-y plane with Th moving out of phase with Ni and C.  The final mode in this complex, which lies in this same frequency range only accidentally, is a C-Th stretch mode polarized along the $\hat z$ axis with a stationary Ni atom, whose dispersionless nature suggests a picture of a local mode for this vibration.\\
$\bullet$
There is a single non-dispersive branch at 25 meV which involves motion of all atoms
along the $c$-axis: Th along $+\hat z$, and both N and C along $-\hat z$. Thus this is the
C-Ni-C trimer moving out of phase with the Th atoms in the Th-C-Ni-C-Th unit.\\
$\bullet$
Four branches in the 35-55 meV range contain one degenerate pair with in-plane C displacements
in the $\hat x$ (and $\hat y$) direction, and another pair with similar frequency
that involve C along +$\hat x$ and Ni along --$\hat x$, and its degenerate partner. Thorium is
involved in only a very minor way in these modes, as anticipated by their high energy.
\\
$\bullet$
The two highest frequency branches around 70 meV involve Ni-C stretch motion of a
relatively strong bond and vibration of the light C atoms.  One has the two C atoms moving
out of phase with
the Ni atom, while for the other the Ni atom is stationary.

As mentioned earlier,
evidence of directional ({\it viz.} covalent) Th-C bonding as well as Ni-C
bonding along the $c$-axis was provided by SI in density contour plots, whereas bonding
within the $a-b$ plane is essentially metallic throughout the structure. This covalent
bonding is reflected in the character of the high frequency modes.

\section{Frozen atom displacements}
\begin{figure}[!htb]
\noindent
\begin{center}
\includegraphics[width=0.5\textwidth,angle=0]{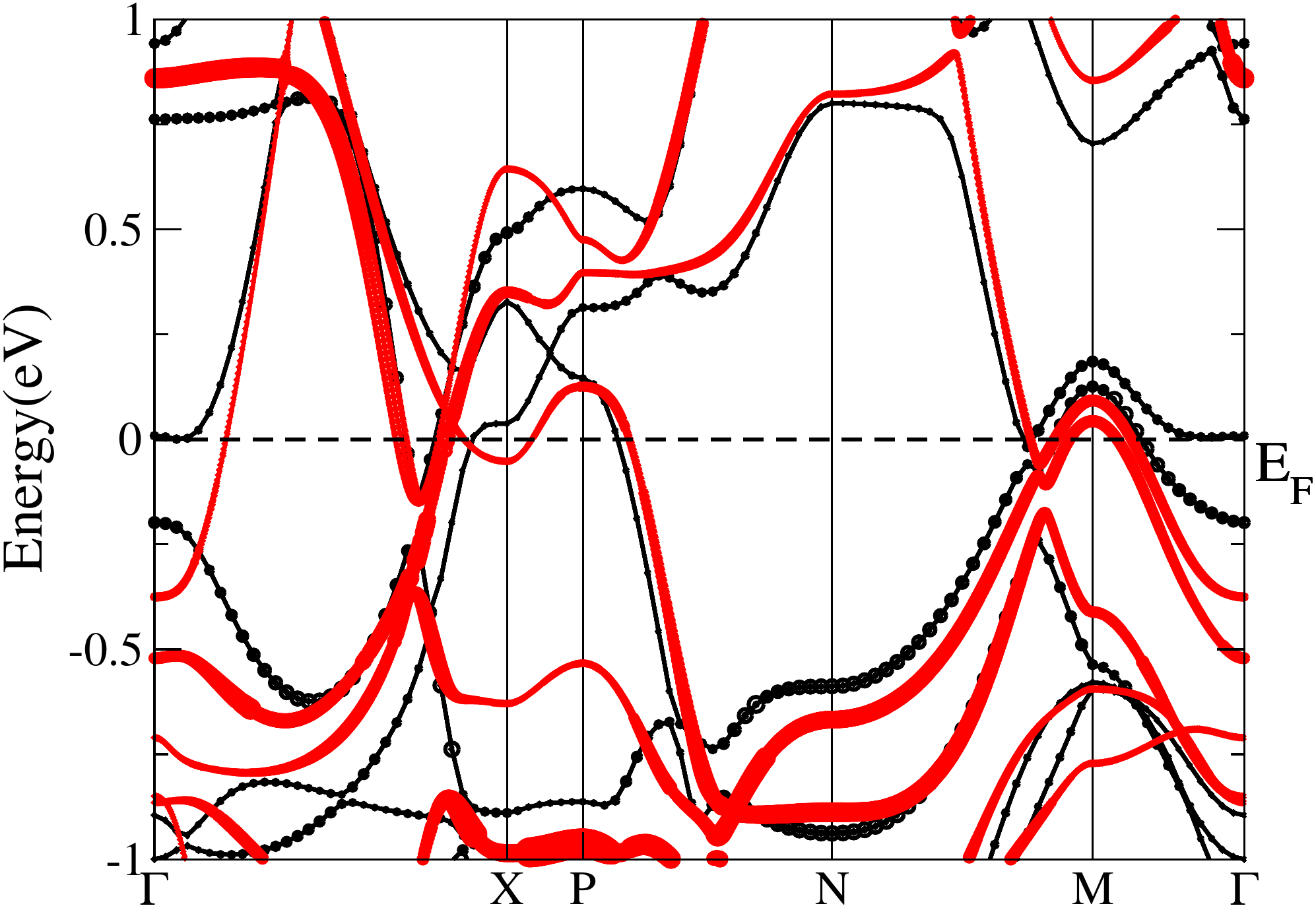}
\end{center}
\caption{(Color online) Bands near E$_F$ before and after C displacement along the stretch mode.
Note that the shift of the flat band at $\Gamma$, more than 2 eV/\AA, is one of the largest.
Black: Ni $3d$ character in Th$_{2}$NiC$_{2}$ with undisplaced C atom,
with a Ni-C distance of 1.9275\AA. Red: Nickel $3d$ character in Th$_{2}$NiC$_2$ with
C atom displaced to a Ni-C bond length of 2.1005\AA. Band shifts near E$_F$ relevant
to electron-phonon coupling are discussed in the text.}
\label{fig:Th2NiC2displaceC}
\end{figure}

To study the sensitivity of individual bands at the Fermi surface to the Ni-C bond
stretch, C atoms were displaced to $z_C$=0.170, a rather large 0.17\AA~displacement
that allows easier identification of the deformation potentials (the rate of change
of band energy with respect to atomic motion, which reflects an important contribution
to electron-phonon matrix elements). The resulting band structure near E$_F$ is
plotted along with that of the equilibrium structure in Fig.~\ref{fig:Th2NiC2displaceC}.
From the Ni fat band character included in Fig.~\ref{fig:Th2NiC2displaceC}, one can
observe that bands with strong Ni character are influenced much less by this C
displacement compared with bands with little Ni character, which have more C and Th character.
The deformation potentials range from nearly zero up to 2 eV/\AA, indicating
significant but not unusually large coupling to Fermi surface states. Note that the largest
coupling is to the zone center flat band at E$_F$.

For some indication of the coupling of Th motion to electronic states, Th was displaced along
the $c$-axis by $ 0.1$\AA, again a large phonon amplitude for a heavy atom.
Even so, the bands
near the Fermi level shift very little compared to C displacement, indicating weak coupling
to Th motion.

\section{Impact of Band Anomalies}
Unusual behavior of a material can often be linked directly to fine structure at E$_F$,
with examples showing new characteristics appearing recently. The most common situation
is that of conventional van Hove singularities (see below), which are generic to a
band structure. More interesting because they are unconventional are other types of
band anomalies.  An example showing some 
similarities to Th$_2$NiC$_2$ is NaAlSi: it is a ternary semimetal with superconducting
T$_c$=7 K, and displays an unusually sharp peak in N(E) at the Fermi level due to 
hybridization between slightly overlapping valence and conduction bands.\cite{naalsi}
Another example is the intermetallic compound NbFe$_2$, which displays a quantum critical
point slightly off stoichiometry.\cite{nbfe2expt}
Study of the band structure led to the prediction that the critical behavior is due to
a van Hove singularity with one divergent effective mass\cite{nbfe2UCD}
(a [$\delta k]^3$ dispersion in one direction) that was confirmed by coherent potential
approximation alloy calculations.\cite{nbfe2PRL}
The enigmatic intermetallic compound TiBe$_2$,
a highly enhanced incipient ferromagnet with strongly temperature dependent susceptibility
and Knight shift, displays DOS fine structure\cite{tibe2} at
E$_F$ as well. 

Experimentally, the interesting features of Th$_2$NiC$_2$ are (1) superconductivity in a
new Ni-based material, (2) T$_c$ is enhanced by 30\% by minor doping by Sc, and (3) the
extrapolation of the heat capacity suggests vanishing electronic DOS at the Fermi level,
causing questions about the observed metallic and superconducting behavior. These
first principles calculations presented above
reveal unusually fine structure in the electronic bands at E$_F$, structure that should
affect several properties of the undoped compound. In this case, the fine structure is
due to spin-orbit coupling induced splitting of band degeneracies together with the
usual mixing of bands. The effective Fermi energy $\varepsilon_F
\sim$ 10 meV implies non-degenerate electron gas behavior beginning at a few tens of Kelvins;
the energy variation of N(E) requires attention when considering thermodynamic and 
transport coefficients. The standard
simple use of ``N(E$_F$)'' has to be replaced with a formalism that incorporates the
non-degeneracy.  Additionally, the Ni-C stretch mode with energy $\omega_{s}$ = 70 meV, 
which seems moderately strongly coupled
to FS states and is a candidate for the pairing mechanism, will have non-adiabatic contributions:
N(E) varies strongly over the regions (-$\omega_{s}$,$\omega_{s}$) so that standard
Eliashberg theory based on the Born-Oppenheimer approximation no longer holds. It should also
be noted that N(E) beyond the narrow peak on the positive energy side is $\sim$50\% larger
than on the negative energy side; electron- and hole-doping should be substantially different,
with electron-doping providing larger N(E$_F$).

Variation of N(E) on a scale of importance for properties, discussed especially for van Hove
singularities, has a long history and large literature. This area of interest was spurred
initially by the flat bands and van Hove singularities that arise in the superconducting
A15 compounds,\cite{carbotte,wep1980,freericks,wepPRL} and figured even more prominently for the high 
temperature superconducting cuprates; see for example Ref. [\onlinecite{norman}]. 
A general formalism for including effects of both thermal smearing and electron-phonon
broadening  was given by one of the present
authors.\cite{wep1982}  At the most fundamental level, the chemical potential (``Fermi level'')
becomes dependent on temperature. For the symmetric peak in N(E) that we obtain, this shift
may be minor; however, off-stoichiometry can readily change this. 
In the normal state the consequences of DOS variation affect the spin
susceptibility and transport properties as well as the electronic heat capacity, and also
complicates the effects due to defects\cite{wep1982} through broadening of N(E).

\section{Summary}
First principles methods have been applied to illuminate the electronic,
vibrational, and doping behavior in the newly discovered superconductor Th$_2$NiC$_2$. While
the Fermi level is found to fall in a rather low density of states region, we obtain a value
of N(E$_F$) that should be easily detectable experimentally, and which we have not been able
to reconcile with the extrapolation of specific heat data from above T$_c$ that implies
an essentially vanishing N(E$_F$). 
The Ni $3d$ bands of this compound are essentially filled, consistent with a ``neutral Ni''
picture, but some regions of the Fermi surface still have considerable Ni character while
other portions are more heavily C-derived.

The distinctive aspect of the electronic structure is very fine structure lying precisely
(but accidentally) at the Fermi level, a result that is only obtained when spin-orbit
coupling is included.  Slight hole doping ($\sim$1\%) can change N(E$_F$) by a factor of
two, and it is plausible that such intrinsic hole doping helps to stabilize the material.
From this reference point, the rise in N(E$_F$) with further doping can help to
account for the observed increase in T$_c$ with hole doping. 
Within this scenario, electron doping would strongly increase N(E$_F$) and thereby the
superconducting T$_c$.

Doping with Sc for Th that increases T$_c$ by 30\% in this 
system has been studied in
two ways: virtual crystal approximation, and substitutional replacement in a supercell.  
While VCA gives a rigid band result as anticipated, the supercell method suggests there
is some shift in spectral density just below the Fermi level due to Th replacement by Sc.
The movement of the Fermi level
downward by this substitution also results in somewhat stronger
Ni-C bonding character, hence larger electron-phonon matrix elements participating to
an increase in T$_c$.
The phonon dispersion curves display distinct clusters of branches distributed up to 70 meV.
The two high frequency branches derive from the C-Ni-C stretch modes.

Frozen phonon calculations involving displacements of (separately) the C and Th atoms
indicates significant electron-phonon coupling of the C atoms but insignificant
coupling to Th motion. The superconductivity in this system seems consistent with
an electron-phonon mechanism as seems to be the case in most other Ni-C superconductors
such as those mentioned in the Introduction, but we do not have a quantitative estimate
of the strength of coupling. 
The phonon calculations do not include Th substitution by Sc, but an interesting possibility
is that the light Sc atom could produce a local-mode vibron and possible unconventional contribution to
electron-phonon coupling.

\section{Acknowledgments}
We thank Z. Fisk for early notification of the experimental data on Th$_2$NiC$_2$.
This work was supported by DOE SciDAC grant DE-FC02-06ER25794 and a collaborative effort with
the Energy Frontier Research Center {\it Center for Emergent Superconductivity}
through SciDAC-e grant
DE-FC02-06ER25777.

\end{document}